\documentstyle[twoside,fleqn,espcrc2,epsfig]{article}

\def\Li(#1,#2){{\rm{Li}}_{#1}(#2)}

\title{Sudakov Logarithms in Four-Fermion Electroweak Processes
       at High Energy}

\author{S. Moch\address{
        Deutsches Elektronen-Synchrotron, DESY, 
        Platanenallee 6, 15738 Zeuthen, Germany}
        }

\begin{document}

\pagestyle{empty}

\begin{abstract}
We discuss recent results for the asymptotic behavior of 
fermion scattering amplitudes in the Sudakov limit. 
The results include next-to-next-to-leading logarithmic electroweak corrections 
and are used for the analysis of fermion-antifermion pair production
in $e^+e^-$ annihilation at high energy. 
The importance of the subleading logarithmic contributions is
emphasized and the fermionic contributions to the Abelian form factor
at two loops are discussed.
\end{abstract}

% typeset front matter (including abstract)
\maketitle

\section{Introduction}
Four-fermion processes are generally considered as
benchmark processes at high energy colliders, with electron
positron annihilation into muon or quark pairs at LEP and the Drell Yan
process at hadron colliders as characteristic examples. 
At energies presently accessible, typically up to 200~GeV, radiative
corrections are dominated by the shift in the $W$ and $Z$ masses as
parametrized by the $\rho$ parameter and by the running of the 
coupling constant. 
Vertex corrections and box diagrams involving gauge bosons are
generally of minor importance. 
In the TeV region, which will be probed at future
colliders like the LHC or TESLA, this picture changes drastically. 
There, double logarithmic corrections become relevant and rapidly
dominant. They were first observed by Sudakov \cite{Sudakov:1956sw} 
in the context of Quantum Electrodynamics for reactions with cuts on
the radiated energy of the photons. 
For electroweak interactions, the large negative corrections arise
from the exchange of gauge bosons which remain uncompensated if one 
restricts the analysis to exclusive final states, consisting for
instance of a fermion antifermion pair only.

Complete one-loop corrections to the four-fermion process are
available since long (see e.g.\ \cite{Beenakker:1991ca}). 
In recent years, the systematic study of double logarithmic
electroweak corrections has received a lot of 
attention~\cite{Ciafaloni:1998xg,Beccaria:1998qe,Kuhn:1999nn,Fadin:1999bq,Beccaria:1999fk,Hori:2000tm,Kuhn:2001hz},
(see \cite{Melles:2001ye} for a review).
While these studies are fairly straightforward for a theory with
massive gauge bosons only, an important complication
arises in the Standard Model due to the presence of massless photons in the final state. 
Events with soft and hard photon radiation are normally included in
the sample, thus a ``semi-inclusive'' definition of the cross
section is closest to the experimental analysis.

At present, all large contributions up to the next-to-next-to-leading
logarithms have been evaluated~\cite{Kuhn:2001hz} for the form factor, 
for four-fermion scattering in a spontaneously broken $SU(2)$ gauge
theory and, last not least, for the Standard Model with
$W$, $Z$-bosons and the massless photon.
In addition, other reactions have been investigated in the high
energy limit, such as gauge boson pair production. There, the next-to-leading
logarithmic terms have been calculated~\cite{Melles:2000gw,Denner:2000jv} 
for the exclusive production of transversly and longitudinally
polarized $W$-bosons. 
But even in inclusive boson fusion 
processes, enhanced electroweak corrections are
present~\cite{Ciafaloni:2000df}, 
a consequence of the lack of compensation between virtual and real
emission due to the non-Abelian charges of the incoming particles.

\section{The Abelian form factor}
Let us begin with a discussion of the form factor of an 
Abelian vector current in the Sudakov limit.
In Born approximation, one writes 
\begin{eqnarray}
{\cal F}_B=\bar\psi(p_2)\gamma_\mu\psi(p_1) ,
\end{eqnarray}
and we study the limit
$s=(p_1-p_2)^2 \to-\infty$ with
on-shell massless fermions, $p_1^2=p_2^2=0$, and
massive gauge bosons, $M^2\ll -s$.
For convenience we choose
$p_{1,2} = (Q/2,0,0,\mp Q/2)$ so that  $2 p_1\!\cdot\! p_2 = Q^2=-s$.

The large logarithmic corrections in the Sudakov limit can be resummed
to all orders of perturbation
theory~\cite{Sen:1981sd,Korchemsky:1989pn}, 
such that the asymptotic behavior of the form factor is obtained from
\begin{eqnarray}
\label{eq:evolsolf}
\lefteqn{
{\cal F}=F_0(\alpha(M^2))\exp \Bigl\{\int_{M^2}^{Q^2}{{\rm d} x\over
  x} }
\\ \lefteqn{ 
\times
\Bigl[\int_{M^2}^{x}{{\rm d} x'\over x'}\gamma(\alpha(x'))+\zeta(\alpha(x))
+\xi(\alpha(M^2))\Bigr]\Bigr\} {\cal F}_B .
} \nonumber
\end{eqnarray}
The resummation exploits the factorization of the scattering amplitude 
in the high energy limit into products of functions $h$, $J$ and $S$, 
which organize the large corrections corresponding to a particular momentum region.
The jet function $J$ contains the full dynamics of collinear 
momentum regions. It includes all leading logarithms. 
The soft function $S$ summarizes the dynamics of soft momenta, 
while the function $h$ describes the short distance dynamics of the hard
scattering process. It contains no large logarithms.
The predictive power of the factorized amplitude then follows from the 
properties of the individual functions. 
In particular, $J$ and $S$ can be defined in terms of operators, 
which obey renormalization group equations with calculable anomalous 
dimensions~\cite{Sen:1981sd,Korchemsky:1989pn,Sterman:1987aj,Magnea:1990zb,Contopanagos:1997nh}.

The next-to-next-to-leading logarithmic corrections include all the terms of the 
form $\alpha^n\log^{2n-m}(Q^2/M^2)$ with $m=0,~1,~2$. 
To this accuracy, one needs for the anomalous 
dimensions $\gamma$,  $\zeta$ and $\xi$ in Eq.~(\ref{eq:evolsolf}) 
the one-loop results 
\begin{eqnarray}
\lefteqn{
\gamma^{(1)} = -2C_F \,,\quad
\zeta^{(1)} = 3C_F\,,\quad
\xi^{(1)} = 0 \,,
}
\nonumber \\
\lefteqn{
F_0^{(1)} = - C_F \left({7\over 2}+{2\over 3}\pi^2 \right) \,,
}
\label{eq:oneloopadim}
\end{eqnarray}
as well as the two-loop result \cite{Kodaira:1982nh} for 
the anomalous dimension $\gamma^{(2)}$, where we define 
$\gamma = \left(\frac{\alpha}{4\pi}\right) \gamma^{(1)} +
\left(\frac{\alpha}{4\pi}\right)^2 \gamma^{(2)} + \ldots$
and similarly for $\zeta$, $\xi$ and $F_0$.

In the $\overline{MS}$-scheme, 
including $n_f$ light fermions and $n_s$ light scalars 
in the fundamental representation, $\gamma^{(2)}$ reads 
\begin{eqnarray}
\lefteqn{
\gamma^{(2)}
\,=\, 
}
\nonumber \\
\lefteqn{
        - 2 C_F \biggl[
          \biggl( 
          {67 \over 9} 
        - {\pi^2 \over 3}  \biggr) C_A 
        - {20 \over 9 } T_F n_f 
        - {8 \over 9 } T_F n_s
       \biggr] ,
}
\label{eq:gamma2}
\end{eqnarray}

It is instructive, to expand Eq.~(\ref{eq:evolsolf}) for the form
factor at two loops up to next-to-next-to-next-to-leading logarithmic (N$^3$LL) accuracy,
\begin{eqnarray}
{\cal F}^{(2)}_{\rm LL}&\!\!\!=\!\!\!&
{1\over 8}{(\gamma^{(1)})^2}\ln^4(z){\cal F}_B ,
\label{eq:llf}\\
{\cal F}^{(2)}_{\rm NLL}&\!\!\!=\!\!\!& 
{1\over 2}\left({1\over 3}\beta_0 - \zeta^{(1)}\right) \gamma^{(1)}
\ln^3(z){\cal F}_B ,
\label{eq:nllf}
\\
{\cal F}^{(2)}_{\rm N^2LL}&\!\!\!=\!\!\!&{1\over 2}\biggl(\gamma^{(2)}+
\left(\zeta^{(1)}-\beta_0\right)\zeta^{(1)}
\nonumber \\ && \qquad
+F_0^{(1)}\gamma^{(1)}\biggr)
\ln^2(z){\cal F}_B ,
\label{eq:nnllf}
\\
{\cal F}^{(2)}_{\rm N^3LL}&\!\!\!=\!\!\!& - \left( \zeta^{(2)} + \xi^{(2)} 
+F_0^{(1)}\zeta^{(1)}\right)
\ln(z){\cal F}_B ,
\label{eq:nnnllf}
\end{eqnarray}
where $z=M^2/Q^2$. 

Employing the results of Eqs.~(\ref{eq:oneloopadim}) and (\ref{eq:gamma2}),
we see a particular pattern of growing coefficients of the logarithms,
which reflects the general structure of logarithmically enhanced
electroweak corrections.
The relatively small coefficient of the leading logarithm and the 
large coefficient of the N$^2$LL term in the 
form factor are clearly indicative of the importance of subleading
logarithmic corrections. 
At N$^3$LL accuracy, the still unknown quantities $\zeta^{(2)}$ and
$\xi^{(2)}$ enter. 

Finally, we mention an equivalent approach to calculate the 
anomalous dimensions in Eq.~(\ref{eq:evolsolf}). 
This is the so-called strategy of regions
\cite{Beneke:1998zp,Smirnov:1998vk,Smirnov:1999bz}, which applies 
expansions in various kinematical regions directly to Feynman diagrams 
contributing to the form factor. From an evaluation of the Feynman 
diagrams to logarithmic accuracy, one may read off the results 
of Eqs.~(\ref{eq:oneloopadim}) and (\ref{eq:gamma2}).

\section{Electroweak four-fermion processes}
Let us now investigate the four-fermion scattering
at fixed angles  in the limit when
all the  invariant energy and momentum transfers
of the process are far larger than the
gauge boson mass,  $|s|\sim |t| \sim |u| \gg M^2$.

For the analysis of the four-fermion amplitude, one has to account 
for the different ``color'' and Lorentz structures, which naturally
induce the dependence on $s$, $t$ and $u$.
A complete  basis consists of four independent chiral amplitudes, each of them
of two possible ``color'' structures.

Let us denote by $\tilde {\cal A}$ the amplitude with
the collinear logarithms  factored out.
It contains only the logarithms of soft nature and 
can be  represented as a vector in the color/chiral basis.
It satisfies an evolution equation which is known from 
QCD~\cite{Sterman:1987aj,Contopanagos:1997nh,Sen:1983bt,Kidonakis:1996aq,Kidonakis:2001nj},
\begin{eqnarray}
{\partial \over \partial \ln{Q^2}}\tilde {\cal A}=
{\bf \chi}(\alpha(Q^2))\tilde {\cal A} \, .
\label{evoleqa}
\end{eqnarray}
Here  $\chi(\alpha)$ is the matrix of the soft anomalous dimensions, 
given for instance in \cite{Kuhn:1999nn,Kidonakis:1996aq,Kidonakis:2001nj},
and the matrix elements of $\chi(\alpha)$ depend on the scattering
angle through $t$ and $u$. 
Upon integration, eq.~(\ref{evoleqa}) resums all next-to-leading logarithms. 

The solution of eq.~(\ref{evoleqa}) reads 
\begin{eqnarray}
\tilde {\cal A}=
\sum_i\tilde {\cal A}_{0i}(\alpha(M^2))\exp\biggl[\int_{M^2}^{Q^2}
{{\rm d} x\over x}\chi_i(\alpha(x))\biggr] ,
\label{eq:evolsola}
\end{eqnarray}
where $\chi_i(\alpha)$ are eigenvalues of $\chi(\alpha)$, 
and $\tilde {\cal A}_{0i}(\alpha)$ 
are $Q$-independent  eigenvectors of $\chi(\alpha)$
which are determined by matching the initial conditions for the evolution
equation at $Q=M$ to a full one-loop calculation to logarithmic accuracy including the
constant terms.

These angular dependent terms as well as those in the starting values of
$\tilde {\cal A}$ discussed below
are important for the calculation of the N$^2$LL terms 
in cross sections and asymmetries.
To illustrate the significance
of the subleading  contributions, 
let us discuss a standard model inspired
example such as the case of the $SU(2)_L$ group
with $n_f = 6$ and one scalar boson.
After integrating the  differential cross section
over all angles one obtains
\begin{eqnarray}
\sigma^{(1)}&\!\!\!=\!\!\!&\biggl[-3\ln^2\left({s\over M^2}\right)
+{80\over 3}\ln\left({s\over M^2}\right)
\nonumber \\ &\!\!\!\!\!\!&
-\left({25\over 9}+3\pi^2\right)\biggr]\sigma_B\,,
\nonumber \\
\sigma^{(2)}&\!\!\!=\!\!\!&\biggl[{9\over 2}\ln^4\left({s\over M^2}\right)
-{449\over 6}\ln^3\left({s\over M^2}\right)
\nonumber \\ &\!\!\!\!\!\!&
+\left({4855\over 9}+{37\over 3}\pi^2\right)\ln^2\left({s\over M^2}\right)
\biggr]\sigma_B\,,
\label{sigsu2s}
\end{eqnarray}
if initial and final state fermions have the same isospin.
Again we observe a relatively small coefficient of the leading
logarithm and the large coefficient of the N$^2$LL term.

To analyze the electroweak correction to the process
$f'\bar f'\rightarrow f\bar f$,
the approximation with $W$ and $Z$ bosons of the same mass  $M$, a Higgs boson
of the mass $M_H\sim M$
and massless quarks and leptons is employed.
A fictitious photon mass $\lambda$ has to be introduced
to regularize the infrared divergences.
The photon is, however, massless and the corresponding infrared
divergent contributions  should be
accompanied by  the real soft photon radiation
integrated to some resolution energy  $\omega_{res}$
to get an infrared safe cross section independent of an auxiliary photon mass.
To study the virtual corrections in the limit of the vanishing
photon mass the general approach of the infrared evolution equations
developed in~\cite{Fadin:1999bq} (see also  references therein)
is particularly useful.

Explicit expressions for the N$^2$LL results are given in \cite{Kuhn:2001hz}.
Numerically, for the reaction $e^+e^-\rightarrow \mu^+\mu^-$
as a typical example, one obtains a correction factor
\begin{eqnarray}
  R^{\mu^+\mu^-} =
    1 - 1.39\,L(s) + 10.12\,\ell(s) - 31.33\,a
\nonumber \\
    + 1.42\,L^2(s) - 18.43\,L(s)\ell(s)
    + 99.89\,\ell^2(s)
\end{eqnarray}
with $a = g^2/16\pi^2 = 2.69 {\cdot} 10^{-3}$,
$\ell(s) = a\,\ln s/M^2$, $L(s) = a\,\ln^2 s/M^2$.
In particular, we note the sizeable coefficient of the N$^2$LL term, 
thus demonstrating the importance of the subleading logarithmic corrections.
To get the infrared safe result for the semi-inclusive cross sections one has to
include the standard QED corrections due to soft photon emission and the pure
QED virtual corrections.

\section{Fermionic contributions at two loops}
In view of the large subleading logarithms, 
it seems desirable to attempt the evaluation of all 
two-loop terms linear in the logarithm or even 
of the two-loop constant terms.

As a first step, the corrections due to
$n_f$ massless fermions and $n_s$ charged massless scalars in the fundamental representation 
have been calculated for the Abelian form factor \cite{BKM}.
Let us define the form factor in terms of scaling functions as 
\begin{eqnarray}
\lefteqn{
{{\cal F} \over {\cal F}_{\rm B}}
= 
        1 
        +       
        {\alpha \over 4 \pi} {\cal F}^{(1)} \!\!
        +
        \! {\alpha^2 \over 16 \pi^2} \! \left(
   n_f  
        {\cal F}^{(2)}_{n_f} 
 + n_s
        {\cal F}^{(2)}_{n_s} \right)\! .\,\,
}
\label{eq:FFnfscaling}
\end{eqnarray}

In the high energy limit, i.e. $z=M^2/Q^2 \to 0$,  
the large logarithms in the scaling functions are 
\begin{eqnarray}
\lefteqn{
{\cal F}^{(1)}  
\,=\, 
         - \ln^2 (z)- 3 \ln(z) - {7 \over 2} - {2 \over 3} \pi^2 \, ,
}
\\
\lefteqn{
{\cal F}^{(2)}_{n_f} 
\,=\, 
       {4 \over 9} \ln^3(z) + {38 \over 9} \ln^2(z) +
                        {34 \over 3} \ln(z)
                        + {115 \over 9} }
\nonumber \\
\lefteqn{
\quad\quad\quad
+ {16 \over 27} \pi^2 \, ,
}
\label{eq:FFlogs-nf}
\\
\lefteqn{
{\cal F}^{(2)}_{n_s} 
\,=\,
 {1 \over 9} \ln^3(z) + {25
          \over 18} \ln^2(z) 
  + {23 \over 6} \ln(z) + {157 \over 36} 
}
\nonumber \\
\lefteqn{
\quad\quad\quad
+ {10 \over 27} \pi^2 \, ,
}
\label{eq:FFlogs-ns}
\end{eqnarray}
where the terms up to $\ln^2(z)$ in Eqs.~(\ref{eq:FFlogs-nf}) and
(\ref{eq:FFlogs-ns}) agree with the result from the evolution equation, 
i.e. Eqs.~(\ref{eq:llf})--(\ref{eq:nnllf}). 
The coefficients of the terms proportional to $\ln(z)$ in Eqs.~(\ref{eq:FFlogs-nf}) and
(\ref{eq:FFlogs-ns}) are a new result \cite{BKM}.
With the help of Eq.~(\ref{eq:nnnllf}), one can determine the
contributions of light fermions and light charged scalars to the sum
of the anomalous dimensions $\zeta^{(2)}$ and $\xi^{(2)}$,
\begin{eqnarray}
\lefteqn{
\zeta^{(2)} + \xi^{(2)} \Biggl|_{n_f+n_s}
\!\!\!\!\! =\, 
        -  C_F \biggl[
          {34 \over 3 } T_F n_f 
        + {23 \over 6 } T_F n_s
       \biggr] .
}
\end{eqnarray}

\begin{figure}[htb]
\begin{center}
\epsfig{file=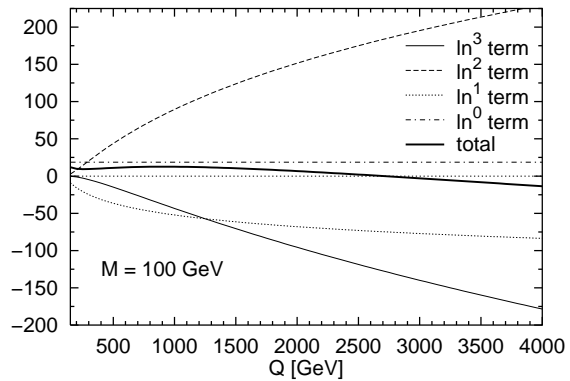,scale=0.975}
\end{center}
\caption{\small 
  The fermionic contribution ${\cal F}^{(2)}_{n_f}$ 
  of the Abelian form factor at two loops. 
  Plotted are the complete result of Eq.~(\ref{eq:FFlogs-nf}), 
  as well as the individual contributions of the large logarithms.}
\label{fig:FFlogs}
\vspace{-0.5cm}
\end{figure}

To study the numerical size of the logarithmic approximation of
Eq.~(\ref{eq:FFlogs-nf}), we plot ${\cal F}^{(2)}_{n_f}$ in Fig.~\ref{fig:FFlogs} 
for a typical mass scale of $M=100$ GeV. 
We show the complete result as calculated in Eq.~(\ref{eq:FFlogs-nf})
and the individual contributions of the large logarithms.
We observe again the pattern of large cancellations between leading and subleading logarithms.

\noindent

Finally, we note that the fermionic contribution to $\gamma^{(3)}$ at 
three loops has recently been calculated \cite{Moch:2002sn,Berger:2002sv}.
Together with the two-loop constant terms in Eq.~(\ref{eq:FFlogs-nf}), 
this result is a step towards the determination of the first five powers of logarithms of
$n_f$-enhanced terms in the Abelian form factor to all orders in perturbation theory.
Furthermore, with the known factorization properties of the full four-fermion amplitude 
\cite{Catani:1998bh,Sterman:2002qn,Glover:2001ev}, 
the result of Eq.~(\ref{eq:FFlogs-nf}) can even be used to get the
complete fermionic contribution at two loops to logarithmic accuracy. 

\section{Conclusions}
We have discussed the asymptotic behavior of the four-fermion
scattering amplitudes in the Sudakov limit. 
We have emphasized the close connection between factorization and
resummation and derived explixcit expressions for the
next-to-next-to-leading logarithmic electroweak corrections 
with the help of evolution equations. 

The subleading logarithmic corrections were found to be sizeable and 
motivated the exact two-loop calculation of the fermionic contribution 
to the Abelian form factor. 

The studies presented here are a first step towards refined precision 
analyses of the high energy asymptotic behavior of four-fermion amplitudes
in non-Abelian gauge models and the Standard Model.

\smallskip
{\bf Acknowledgments}\\
We acknowledge gratefully pleasant collaborations 
with B.~Feucht, J.H.~K\"uhn, A.A.~Penin and
V.A.~Smirnov.

\end{document}